\documentclass[12pt,aps,preprint,prl,amsmath,amssymb]{revtex4}

\usepackage{epsfig}

\def\da{^\dagger}

\newcommand{\Op}[1]{{\boldsymbol{\mathrm{\hat{#1}}}}}

\begin{document}

\title{Molecular Quantum Computing by an Optimal Control 
Algorithm for Unitary Transformations}

\author{Jos\'e P. Palao $^{(a,b)}$ and Ronnie Kosloff $^{(a)}$}

\affiliation{$^{(a)}$ Department of Physical Chemistry and the Fritz 
Haber Research Center for Molecular Dynamics, Hebrew University, 
Jerusalem 91904, Israel\\
$^{(b)}$ Departamento de F\'{\i}sica Fundamental II, Universidad de 
La Laguna, La Laguna 38204,Spain}



\begin{abstract}
\noindent

Quantum computation is based on  implementing 
selected unitary transformations which represent algorithms.
A generalized  optimal control theory is used to 
find the driving field that generates a prespecified unitary transformation.
The approach is illustrated in the implementation of one and
two qubits gates in model  molecular systems. 
\\

\noindent{PACS number(s): 82.53.Kp 33.90.+h 32.80.Qk 03.67.LX}
\\

\end{abstract}

\maketitle



A universal model of a quantum computer  can be constructed from
an array of two level systems used as registers (qubits). 
Any general quantum gate can be decomposed to a one and two qubit
unitary transformation \cite{nielsen00}.
Thus a physical realization of a quantum computer
should be able  control such unitary transformations
by use of external driving fields.
Realizations based on nuclear 
magnetic resonance techniques already have been demonstrated for  non-trivial
quantum gates. For example,  a three qubit quantum Fourier 
transform has been performed by a sequence of pulses
corresponding to one and two qubit transformations
\cite{weinstein01}. This implementation was based 
on addressing each qubit independently 
by their spectral separation.
However, when many qubits become involved in a computation, the spectrum
becomes congested so that it becomes more difficult 
to address each qubit individually. 
The total fidelity of the algorithm
will depend on the accumulation of errors at each step.
Another difficulty results from decoherence processes which are
unavoidable due to coupling to the external environment. 
These processes will degrade the performance in proportion
to the time required to carry out the computation task.
Thus it is desirable to minimize the number of computation
steps and the total computation time.

An alternative possibility for implementing quantum computing
is the use of molecules driven  by shaped light pulses.
A hypothetical model can be based on using
the vibrational and rotational states as registers.
A shaped light pulse can write data as amplitudes on these states.
The computing algorithm is then implemented by performing
a unitary transformation  employing a second shaped pulse.
In the end  the output can be read by a probe pulse.
This approach implies changing the computation model by tying together 
many single and two qubit operations into one 
resulting in a  combined unitary transformation.

Such an  approach has been used for the experimental 
implementation of elements of coherent computation
in Li$_2$ \cite{vala02,amitay02}. For example, two 
states of the ro-vibrational manifold of an electronic state can form 
a single qubit. By taking advantage of multiple electronic states, 
the evolution of the molecular 
system can be controlled using electromagnetic fields 
in the optical region realized by shaped laser pulses.  
The fast development of pulse shaping technology
can make possible the implementation of simple quantum
gates in molecules. This program has to overcome 
interference from the large number of other molecular 
levels coupled to the field but  not assigned to the qubits. 
The task becomes therefore to implement the quantum algorithm on
the molecular levels used as registers while avoiding the
intervention of other states from the same system.

Tesh et. al. \cite{tesch01},  proposed the use of optimal control theory
(OCT) to calculate  the field which can induce a specific transformation
used for quantum computation. 
Originally, OCT was designed as a method to obtain a
light field which could induce a specific state to state transformation \cite{pierce88,kosloff89}.
In the quantum computing context, the goal is to 
obtain the optimal pulse that induces a 
given unitary transformation, irrespective of the 
initial state of the system. 
The present approach  generalized OCT to obtain directly  
the driving field that induces
a target unitary transformation in the system. The approach
is based on the  equation of motion of the unitary
transformation by making use of the main ingredients of OCT. This  allows the  
utilization of  the large number of tools that have been developed in this field.


The model system consists of a free Hamiltonian $\Op{H}_0$ controlled
by an external field $\epsilon(t)$,
\begin{equation}\label{eq:hamiltonian}
\Op{H}(t)=\Op{H}_0-\Op{\mu}\,\epsilon(t)\,.
\end{equation}
where $\Op{\mu}$ is a system operator. In the  molecular
system, $\Op{\mu}$ is the transition dipole operator and 
$\epsilon(t)$ describes a shaped short light pulse.
For simplicity,  the field $\epsilon(t)$ is assumed real
but the generalization to an electromagnetic field with two
independently controlled polarizations \cite{gerber2001}
is straightforward.

The algorithm is represented in the target time $T$  
by a unitary transformation $\Op{U}(T)$ generated from 
the Hamiltonian Eq. (\ref{eq:hamiltonian}); 
\begin{equation}\label{eq:evolution}
\frac{\partial\Op{U}(t)}{\partial t} = 
-\frac{i}{\hbar}\Op{H}(t)\,\Op{U}(t)\,,
\end{equation}
with the initial condition $\Op{U}(t=0)=\openone$, where $\openone$
denotes the identity operator. The objective is to obtain the optimal
driving field $\epsilon(t)$ that induces a given unitary transformation
$\Op{O}$ at $t=T$, i.e., $\Op{U}(T)=e^{i\phi}\,\Op{O}$.
$\phi$ denotes a physically irrelevant global 
phase that is related to the energy origin and is therefore 
uncontrollable by the field. 
 
The task  is a typical inversion problem which can be solved
by employing a variational procedure maximizing the projection
of the generated  operator on the target operator,
\begin{equation}\label{eq:functional0}
|\tau| = |{\rm Tr}\{\Op{O}\da \Op{U}(T)\}|\,,
\end{equation}
where the projection $(\Op{A} \cdot \Op{B} )$ is defined by 
${\rm Tr}\{\Op{A}\da \Op{B} \}$. The functional $\tau $ is a complex number 
inside the circle $|\tau|\leq N_H$,
where $N_H$ is the dimension of the Hilbert space of the system. 
Equality is reached only when the argument 
of the trace is $e^{i\phi}\,\openone$ 
and then a maximum of $|\tau|$ is equivalent to $\Op{U}(T)=e^{i\phi}\Op{O}$.
The optimal solution is then found by maximizing 
the functional Eq. (\ref{eq:functional0}),
with respect to the control field $\epsilon(t)$.
Since a direct algorithm to maximize $|\tau|$ was not found, a
working alternative is used based on formulating the problem as
the optimization of ${\rm Re}[\tau]$, or of ${\rm Im}[\tau]$, 
or of a linear combination of both.
For simplicity, the optimization of the real part represented by the 
functional $J={\rm Re}[{\rm Tr}\{\Op{O}\da \Op{U}(T)\}]$ is considered.
Two constraints are introduced \cite{pierce88,kosloff89}, the first
restricts the dynamics to obey the Schr\"odinger equation
Eq. (\ref{eq:evolution}), and the second restricts the total field 
energy. Using Lagrange multipliers a modified functional is obtained,
\begin{equation}\label{eq:functional1}
\bar{J}=
{\rm Re}\left[\tau
-\int_{0}^{T}
{\rm Tr}\left\{\left(\frac{\partial \Op{U}(t)}{\partial t}+
\frac{i}{\hbar}\Op{H}(t)\,\Op{U}(t)\right)\Op{B}(t)\right\} dt\right]  
-\lambda \int_{0}^{T}\frac{1}{s(t)}|\epsilon(t)|^2 dt\,,
\end{equation}
where $\Op{B}(t)$ is an operator Lagrange multiplier, $\lambda$ is 
a scalar Lagrange multiplier and $s(t)$ is a shape function
which turns the pulse on and off  \cite{sundermann99}. 
The use of more elaborate
constraints and choices of $\lambda$, allows a higher degree of
control on the shape of the optimal pulse \cite{hornung01,hornung02}. 

Applying the calculus of variations, $\delta \bar{J}=0$ with
respect to $\Op{B}$, $\Op{U}$, and $\epsilon$, a set of equations is 
obtained: a) The Schr\"odinger equation
Eq. (\ref{eq:evolution})  with the initial condition
$\Op{U}(t=0)=\openone$ for $\Op{U}$; b) The inverse Schr\"odinger equation
\begin{equation}\label{eq:bevolution}
\frac{\partial\Op{B}(t)}{\partial t} = 
\frac{i}{\hbar}\Op{B}(t)\,\Op{H}(t)\,
\end{equation}
with the condition $\Op{B}(t=T)=\Op{O}\da$ for $\Op{B}$; 
c) The field equation:
\begin{equation}\label{eq:optimalf1}
{\epsilon(t)} = -\frac{s(t)}{2\,\lambda\,\hbar}\,\,{\rm Im}[\,{\rm Tr}
\{\Op{B}(t)\,\Op{\mu}\,\Op{U}(t)\}\,]\,,
\end{equation}
%


Eq. (\ref{eq:evolution}) and (\ref{eq:bevolution})
represent two counter currents with information from 
the initial condition and the target unitary 
transformation respectively. 
The equations are solved iteratively, the Krotov method 
\cite{tannor92}, similar to the methods
described in Ref. \cite{zhu98}, was found to be the most efficient.
The input is a ``guess'' field,
$\epsilon^{(0)}(t)$, so that in the $k$ iteration ($k=1,2...$): 
(i) $\Op{B}^{(k-1)}(t)$ is propagated backwards from $t=T$ to 
$t=0$ using Eq. (\ref{eq:bevolution}) and $\epsilon^{(k-1)}$;
and (ii) $\Op{U}^{(k)}(t)$ is propagated forward
using Eq. (\ref{eq:evolution}) and $\epsilon^{(k)}$ 
is evaluated using
\begin{equation}\label{eq:optimalf2}
{\epsilon^{(k)}(t)} = 
-\frac{s(t)}{2\,\lambda\,\hbar}\,\,{\rm Im}[\,{\rm Tr}
\{\Op{B}^{(k-1)}(t)\,\Op{\mu}\,\Op{U}^{(k)}(t)\}\,]\,.
\end{equation}
The procedure is repeated until the desired convergence has been 
reached. The hard numerical task is the propagation of the 
operators $\Op{U}$ and $\Op{B}$ with the time dependent 
Hamiltonian for which  a second order Newton polynomial 
integrator \cite{kosloff94} was used.


A direct use of Eq. (\ref{eq:optimalf2}) in the 
algorithm leads to saturation. This is  because the constraint 
related with the field energy becomes more important 
than the original objective.
A remedy is to interpret the right hand side of
Eq.(\ref{eq:optimalf2}), denoted by $\Delta\epsilon^{(k)}$, 
as a correction of the field in the previous interaction 
\cite{bartana97}. The field after the $k$ iteration is then
given by $\epsilon^{(k)}(t)=\epsilon^{(k-1)}(t)+\Delta\epsilon^{(k)}(t)$.
The OCT procedure was first applied to a 1 qubit operation 
the Hadamard rotation (Eq. \ref{eq:hamhad3}).
In a similar fashion,  2-qubit operation such as the 
``controlled not'' were obtained.
The optimal fields generating unitary operations with eight (3-qubits) 
were also attempted. 

It was found that the number of iterations required to converge
the results increased approximately exponentially (or factorially)
with the number of states in the problem. 
A possible reason for this scaling is that not only a specific state to
state transition has to be forced but this has to be carried out
without disturbing the other state to state transitions in $\Op{U}$.
As a result the scaling becomes $O(n!)$ which is consistent with 
numerical experience. 


In a molecular environment, obtaining the optimal field
to carry out an algorithm is more involved. 
The register levels which are used to write the input and output
are only part of a much larger manifold of molecular energy levels.
Considering the advances in pulse shaping techniques
in the visible region of the spectrum the transitions of choice
are electronic.
For such a molecular construction, imposing a unitary 
transformation on the total set of levels which are addressed by the field
is too restrictive.  Relying on the experience that the convergence is
close to factorial, an extremely large number of
iterations would be required to converge. 

The strategy is therefore to restrict the target objective to only
the states used directly as registers keeping the
other states in the system as passive observers.
The reduced objective is obtained by changing  the previous 
expressions ${\rm Tr}\{\Op{A}\Op{B}\}$ to 
$\sum_i \langle i|_R\Op{A}\Op{B}|i\rangle_R\,$,
where $\{|i\rangle_R\}$ is a basis of the subspace of registers.  
The subindex $R$ is used to denote the operators in that subspace, for
example $\Op{O}_R$ denotes the target unitary transformation.
The substitution of the restricted condition  
instead of the trace in  Eq. (\ref{eq:functional0}) keeps 
the maximum condition when the unitary transformation 
$\Op{U}_R(T)$ is equal to 
$\Op{O}_R$. The maximum value becomes  equal to  
the dimension of the subspace $N_R$. 
In the condition for $\Op{B}$ at time $T$, a particular 
dependence must be specified for all the levels. 
For simplicity the identity in the passive subspace has been  chosen. 

\begin{figure}[h]
\vspace{1.2cm}
\hspace{0.16\textwidth}
\psfig{figure=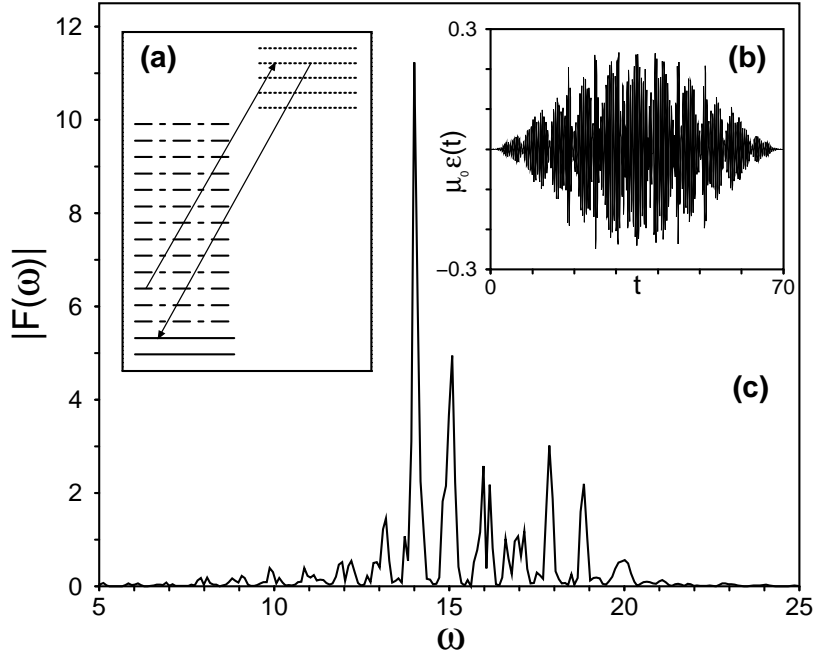,width=0.65\textwidth}
\vspace{.3cm}
\caption{(a) Schematic representation for executing the 
Hadamard rotation. 
Units are chosen such $\hbar=1$.
The solid lines represent the two levels associated with 
the qubit.  
The levels in the ground surface are equally spaced with 
$\omega_g=1$ (for Li$_2$ in the A state $\omega_g=255~ cm^{-1}$) 
and in the excited surface with $\omega_e=0.9$ 
(for Li$_2$ in the E state $\omega_e=241~ cm^{-1}$).
The other model parameters are $\Omega=15$ 
(for Li$_2$ the A to E transitions $\Omega=12400~ cm^{-1}$)
and $\mu_0=0.1$.
The arrows are two
of the possible transitions induced by the driving field.
(b) Optimized field that 
induces a Hadamard rotation in the qubit at $T=70$ (for Li$_2$ $T \sim 1.5$ psec).
(c) Fourier transform of the field .} 
\label{fig:spehad}
\end{figure}

The approach is illustrated using two models. For the first model,
the implementation of  a unitary transformation in one
qubit while minimizing the population transfer to other levels
in the molecule is studied. 
The model consists of a molecule with two electronic surfaces 
described by the Hamiltonian,
\begin{equation}\label{eq:hamhad}
\Op{H}=\left(
\begin{array}{cc}
\Op{H}_g & -\Op{\mu}\epsilon(t) \\
-\Op{\mu}\epsilon(t)& \Op{H}_e 
\end{array}\right)\,,
\end{equation}
where $\Op{H}_g$ and $\Op{H}_e$ are the ground and
excited surface Hamiltonians. The electronic surfaces
are coupled by the transition dipole operator $\Op{\mu}$, 
controlled by the shaped field $\epsilon(t)$. 
The two first levels of the ground electronic surface 
are chosen as the registers representing the qubit.
The model includes $15$ rovibroinc levels
in the ground electronic state and $5$ in the 
excited state described by the  
Hamiltonians, 
\begin{equation}\label{eq:hamhad0}
\Op{H}_g=\sum_{i=1}^{15} E_{gi} 
|g_i\rangle \langle g_i|\,;\;\;\;\;\;\;
\Op{H}_e=\sum_{j=1}^{5} E_{ej} |e_j\rangle \langle e_j|\,.
\end{equation}
shown in Fig. \ref{fig:spehad}.
Next, a transition dipole  operator with equal coupling strength 
between levels was chosen,
$\Op{\mu}=\mu_0 \sum_{i=1}^{15}\sum_{j=1}^{5}
|g_i\rangle \langle e_j|+ {\rm H.c.}\,$,
where ${\rm H.c.}$ denotes the Hermitian conjugate.

In Fig. \ref{fig:spehad} the optimized field is shown 
when the target unitary transformation is a 
Hadamard rotation given by,
\begin{equation}\label{eq:hamhad3}
\Op{O}_{R}^{H} = \frac{1}{\sqrt{2}}
\left(
\begin{array}{cc}
1 & 1 \\
1 & -1 
\end{array}\right)\,,
\end{equation}
restricted to the $2\times 2$ qubit subspace.
The shape function was chosen as $s(t)=\sin^2(2\pi t/T)$ 
and the initial guess
was $\epsilon^{(0)}(t)=s(t)\cos(\Omega t)$, where $\Omega$ is the
frequency of the 00-line of the electronic transition. 
Fig. \ref{fig:spehad} shows that the dominant frequencies
are the ones related to the known vibronic molecular transitions. 
In the time domain the field is split into a symmetric sequence
of sub pulses. A  phase
relation correlating the dominant frequencies is observed in a Wigner plot
(not shown). 
These phase relations guarantee that no population is lost 
to the excited levels.

\begin{figure}[h]
\vspace{1.2cm}
\hspace{0.16\textwidth}
\psfig{figure=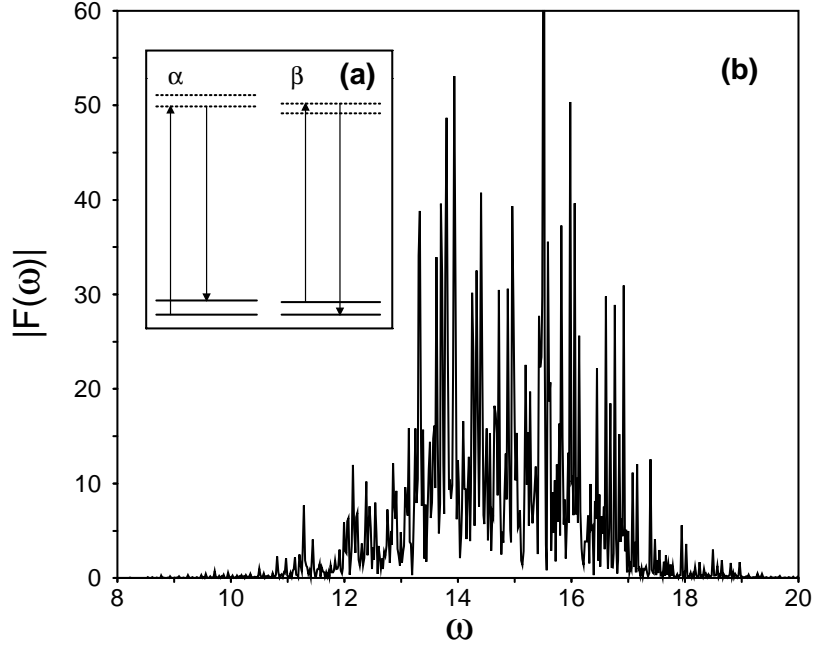,width=0.65\textwidth}
\vspace{.3cm}
\caption{(a) Schematic representation of the execution
of a 2-qubit Fourier transform. The solid
lines represent the levels associated with the qubits.
The frequencies are 
$E_{g\alpha_0}=0$,
$E_{g\alpha_1}=1$,
$E_{e\alpha_0}=15$,
$E_{e\alpha_1}=15.8$,
$E_{g\beta_0}=0$,
$E_{g\beta_1}=0.9$,
$E_{e\beta_0}=14.5$, and
$E_{e\beta_1}=15.2$.
The arrows are examples of transitions induced
by the driving field in each mode. The other model
parameters are $\mu_{0\alpha}=0.1$, $\mu_{0\beta}=0.08$ and 
$\delta_{\alpha\beta}=0.21$. 
(b) Fourier transform of the optimal field that induces
$\Op{O}^{FT}_R$ at $T=320$.}

\label{fig:speft}
\end{figure}

In the second model the two electronic
surfaces, Eq. (\ref{eq:hamhad}), have
two orthogonal vibrational modes, denoted as $\alpha$ and $\beta$.
The first two levels of each vibrational mode in the 
ground electronic surface are chosen as the the physical 
implementation of the two qubits. The goal is 
to induce an operation that involves entanglement 
between them. In this model each vibrational mode of
the ground surface is coupled by the field to the
corresponding mode in the excited electronic surface.
The modes in the excited surface are coupled 
by a static term modeling  Duschinsky rotation. 
This last term which is not controlled by the field 
can generate entanglement between the qubits.
The model consists of two levels for each vibrational
normal mode denoted by $\alpha$ and $\beta$. 
The electronic surface Hamiltonians are
$\Op{H}_g=\Op{H}_{g \alpha}\otimes\openone_{\beta}+\openone_{\alpha}\otimes
\Op{H}_{g \beta}$ and
$\Op{H}_e=\Op{H}_{e \alpha}\otimes\openone_{\beta}+\openone_{\alpha}\otimes
\Op{H}_{e \beta}+\Op{V}_{\alpha \beta}$ with,
\begin{eqnarray}
\Op{H}_{g \nu}&=&\left(
E_{g \nu_0} |g_{0}\rangle_{\nu} \langle g_{0}|_{\nu} +
E_{g \nu_1} |g_{1}\rangle_{\nu} \langle g_1|_{\nu}\right)\,,\nonumber\\
\Op{H}_{e \nu}&=&\left(
E_{e \nu_0} |e_{0}\rangle_{\nu} \langle e_{0}|_{\nu} +
E_{e \nu_1} |e_{1}\rangle_{\nu} \langle e_{1}|_{\nu}\right)\,,\nonumber\\
\Op{V}_{\alpha \beta}&=&\delta_{ \alpha \beta}
\left(|e_{0}\rangle_{\alpha}\otimes|e_{1}\rangle_{\beta} 
\langle e_{1}|_{\alpha}\otimes\langle e_{0}|_{\beta}\,+\, 
|e_{1}\rangle_{\alpha}\otimes|e_{0}\rangle_{\beta} 
\langle e_{0}|_{\alpha}\otimes\langle e_{1}|_{\beta}\right)\,,
\end{eqnarray}
with  $\Op{V}_{\alpha \beta}$ the   Duschinsky 
term that couples the vibrational modes in the excited surface.
A schematic representation of the levels
is given in Fig. \ref{fig:speft}.  
The transition dipole operator is chosen as
$\Op{\mu}=\Op{\mu_{\alpha}}\otimes\openone_{\beta}+
\openone_{\alpha}\otimes\Op{\mu_{\beta}}$,
with
$\Op{\mu}_{\nu}= \mu_{0\nu}
(|e_{0}\rangle_{\nu} + |e_{1}\rangle_{\nu}) 
(\langle g_{0}|_{\nu} + \langle g_{1}|_{\nu})
+ {\rm H.c.}\,$
Operators in the combined Hilbert space  of this system 
are represented by $16\times 16$ matrices.

The target unitary transformation is
a two qubits quantum Fourier transform \cite{weinstein01},
\begin{equation}\label{eq:ft}
\Op{O}_{R}^{FT} = \frac{1}{2}
\left(\begin{array}{cccc}
1 & 1 & 1 & 1 \\
1 & i & -1 & -i \\
1 & -1 & 1 & -1 \\
1 & -i & -1 & i
\end{array}\right)
\end{equation}
where the operator is represented in the basis 
$|g_i\rangle_{\alpha}\otimes|g_j\rangle_{\beta}$. Notice that
operators in the subspace of interest are represented 
by $4\times4$ matrices. The spectra of the field resulting
from the optimization is shown in Fig. \ref{fig:speft}.
The frequencies are confined in the region of the 
vibronic transitions. The  structure of the spectra is much more 
complex than in the previous model. 
An obvious reason is the larger dimension of the subspace 
of interest. In addition the complicated mechanism
required to create entanglement is indirect.
During the pulse significant population is transfered to 
the excited electronic states.
The resulting more complex field  is reflected by 
also in the  Wigner distribution,  where the  field
shows a very high degree of correlation between time and
frequency. 
 

In conclusion, Optimal Control Theory has been generalized  to obtain 
the driving field that generates any target unitary 
transformation. 
In principle the scheme allows the implementation of any 
quantum gate using a single step. 
Compared to the implementation of the gate
using a sequence of known simple pulses, the loss by the 
possible complexity of the shaped pulse can be more 
than compensated by the faster implementation.
Convergence is efficient for a small number of levels.
Besides, with the right choice of the guess field,
the method can obtain the driving field for more 
complex systems where a large number of levels
are involved and the pulses for simple 
operations are unknown, as in molecular systems. 
The advantage of such systems is the short time in which
these algorithms can be executed. Using Li$_2$ as an example,
the Hadamard rotation can be executed in $\sim$ 1.5 psec.
The 2-qubit Fourier transform could be executed in a molecule
for example OCS in approximately $\sim 6$ psec. These fast timescales
give hope that the quantum computation can be carried out
before decoherence processes take place.

J.P.P. acknowledges the Golda Maier Fund of the Hebrew 
University. This work was supported by the Spanish 
MCT BMF2001-3349 and the Israel Science Foundation. 
The Fritz Haber Center is supported by the Minerva 
Gesellschaft f\"ur die Forschung, GmbH M\"unchen, Germany.
The authors thank David Tannor and Zohar Amity for helpful discussions.



\end{document}